%% file: Gault-Paper1.tex
\documentclass[twocolumn]{aastex62}

\newcommand{\ed}[1]{#1}
\graphicspath{{./}{figures/}}

\accepted{March 13, 2019}
\submitjournal{ApJL}

\begin{document}

\title{The Sporadic Activity of (6478) Gault: A YORP-driven event?}

\author[0000-0002-4734-8878]{Jan T. Kleyna}
\affiliation{Institute for Astronomy, University of Hawaii, 2680 Woodlawn Drive, Honolulu, HI 96822, USA}

\author[0000-0001-6952-9349]{Olivier R. Hainaut}
\affiliation{European Southern Observatory, Karl-Schwarzschild-Strasse 2, D-85748 Garching bei M\"unchen, Germany}

\author[0000-0002-2058-5670]{Karen J. Meech}
\affiliation{Institute for Astronomy, University of Hawaii, 2680 Woodlawn Drive, Honolulu, HI 96822, USA}

\author[0000-0001-7225-9271]{Henry H. Hsieh}
\affiliation{Planetary Science Institute, 1700 East Fort Lowell Rd., Suite 106, Tucson, AZ 85719, USA}
\affiliation{Institute of Astronomy and Astrophysics, Academia Sinica, P.O.\ Box 23-141, Taipei 10617, Taiwan}

\author[0000-0003-0250-9911]{Alan Fitzsimmons}
\affiliation{Astrophysics Research Centre, Queen's University Belfast, Belfast BT7 1NN, UK}

\author[0000-0001-7895-8209]{Marco Micheli}
\affiliation{ESA NEO Coordination Centre, Largo Galileo Galilei, 1, 00044 Frascati (RM), Italy}
\affiliation{INAF - Osservatorio Astronomico di Roma, Via Frascati, 33, 00040 Monte Porzio Catone (RM), Italy}

\author[0000-0002-2021-1863]{Jacqueline V. Keane}
\affiliation{Institute for Astronomy, University of Hawaii, 2680 Woodlawn Drive, Honolulu, HI 96822, USA}

\author[0000-0002-7034-148X]{Larry Denneau}
\affiliation{Institute for Astronomy, University of Hawaii, 2680 Woodlawn Drive, Honolulu, HI 96822, USA}

\author[0000-0003-2858-9657]{John Tonry}
\affiliation{Institute for Astronomy, University of Hawaii, 2680 Woodlawn Drive, Honolulu, HI 96822, USA}

\author[0000-0003-3313-4921]{Aren Heinze}
\affiliation{Institute for Astronomy, University of Hawaii, 2680 Woodlawn Drive, Honolulu, HI 96822, USA}

\author[0000-0003-0174-3829]{Bhuwan C. Bhatt}
\affiliation{Indian Institute for Astrophysics, II Block, Koramangala, Bangalore 560 034, India}

\author[0000-0001-9701-4625]{Devendra K. Sahu}
\affiliation{Indian Institute for Astrophysics, II Block, Koramangala, Bangalore 560 034, India}

\author[0000-0001-8690-3507]{Detlef Koschny}
\affiliation{ESA NEO Coordination Centre, Largo Galileo Galilei, 1, 00044 Frascati (RM), Italy}

\affiliation{ESTEC, European Space Agency, Keplerlaan 1, 2201 AZ, Noordwijk, The Netherlands}
\affiliation{Chair of Astronautics, Technical University of Munich, Boltzmannstra\ss e 15, 85748 Garching bei M\"unchen, Germany}

\author[0000-0001-9535-3199]{Ken W. Smith}
\affiliation{Astrophysics Research Centre, Queen's University Belfast, Belfast BT7 1NN, UK}

\author[0000-0001-8429-2739]{Harald Ebeling}
\affiliation{Institute for Astronomy, University of Hawaii, 2680 Woodlawn Drive, Honolulu, HI 96822, USA}

\author[0000-0002-0439-9341]{Robert Weryk}
\affiliation{Institute for Astronomy, University of Hawaii, 2680 Woodlawn Drive, Honolulu, HI 96822, USA}

\author[0000-0002-1050-4056]{Heather Flewelling}
\affiliation{Institute for Astronomy, University of Hawaii, 2680 Woodlawn Drive, Honolulu, HI 96822, USA}

\author[0000-0002-1341-0952]{Richard J. Wainscoat}
\affiliation{Institute for Astronomy, University of Hawaii, 2680 Woodlawn Drive, Honolulu, HI 96822, USA}

\begin{abstract}
On 2019 January 5 a  streamer associated with the 4--10 km main-belt asteroid (6478)~Gault was detected by the ATLAS sky survey, a rare discovery of activity around a  main-belt asteroid. Archival data from ATLAS and Pan-STARRS1 show the  trail in early December 2018, but not  between 2010 and January 2018. The feature has significantly changed over one month, perfectly matching predictions of pure dust dynamical evolution and changes in observing geometry for a short release of dust around 2018 October 28. Follow-up observations with HST show a second narrow trail corresponding to a  brief release of dust on 2018 December 30. Both  releases occurred with negligible velocity. We find the dust grains to be fairly large, with power-law size distributions in the $10^{-5} - 10^{-3}$~m range and  power-law indices of $\sim -1.5$.  Three runs of ground-based data find a signature of $\sim 2\,\rm h$ rotation, close to the rotational limit, suggesting that the activity is the result of landslides or reconfigurations  after YORP spin-up.\\

\end{abstract}


\section{Introduction -- Active Asteroids}
\label{sec:intro}

Active asteroids are objects that have semi-major axes  smaller than Jupiter's, are  orbitally decoupled from Jupiter (with Tisserand parameter  $T_J>3.0$), and exhibit comet-like mass loss \citep{jewitt2015_AstsIV}. They are dynamically distinct from classical comets and have long been present in the outer asteroid belt \citep{kresak1980,levison2006}. Many mechanisms have been proposed to explain the dust observed around active asteroids, including rotational spin up \citep[YORP; e.g.][]{VotBott_YORP_AstIV, Bottke2006}, asteroid impact, collisional debris fields, and sublimation of subsurface ices in main belt comets (MBCs) \citep{jewitt2015_AstsIV}.   Active asteroids offer insight into a range of solar system phenomena (primordial volatiles from MBC sublimation, material composition from rotation and impacts), and it is crucial to study each specimen in detail to determine its mechanism of activity.

\section{A new active asteroid: (6478) Gault}
 
\noindent
The  Hawai`i ATLAS survey   \citep{tonry2018} detected a tail (Fig.~\ref{fig:images}) on asteroid (6478) Gault in images obtained on 2019 January 5, when the object was at a heliocentric distance of $r = 2.48\,\rm au$ \citep{CBET4594}.  A median-combined stack of seven $30\,\rm s$ exposures shows a $135\arcsec$-long tail at PA=290$^{\circ}$.  The ATLAS archive shows that Gault was active  on 2018 December 8, with a $30\arcsec$ tail at PA=290$^{\circ}$. However, we find no evidence of a tail  in previous ATLAS images obtained during 2018 January (it was not observed by ATLAS or Pan-STARRS from 2018 January through December because of its low solar elongation). 
\ed{\cite{ATEL12450} find evidence of the onset, reporting Zwicky Transient Facility (ZTF) archive data  showing significant brightening before 2018 October 31}.

The orbital elements ($e$=0.194, $a$=2.305 au, $i=22.8^{\circ}$, having $T_J$=3.461) are consistent with it being an MBC, albeit with a small semi-major axis.  Discovered in 1988, Gault has an absolute magnitude of 14.4 in the $V$ band, based on $\sim$1000  survey observations, implying a diameter of $\sim 9\,\rm km$, assuming a 4\% geometric albedo typical of comets, or 4 km for a 20\% albedo, representative of asteroids; \ed{we will assume a 20\% albedo unless otherwise stated.}  Gault's tail was also seen in Pan-STARRS1 survey images from 2018 December 17 but appeared stellar in all other images from 2010 September 6 through 2018 January 11, implying that something  in late 2018 triggered the activity.   \ed{A second emission  event detected in mid-January in our data, and reported by \cite{ATEL12450} and \cite{CBET4606}, shows similarities to the episodic activity of 311P, attributed to landslides caused by rotational instability \citep{jewitt2018_311P}.}


\section{Follow-up observations\label{sec:obs}}

\input{ObsTable.tex}

\noindent
Images from the Canada-France-Hawai`i (CFHT) 3.6 m telescope on 2019 January 6 show a tail $\gtrsim 4.3\times 10^5\,\rm km$ long.   The central  brightness measured with both  ATLAS and CFHT  is more than a magnitude brighter than predicted by Jet Propulsion Laboratories (JPL) Horizons, implying  significant excess material within the ground-based seeing disk.  Other groups  also reported an extended tail \citep{CBET4597}.  Both a set of 17 $120\,\rm s$ SDSS-$g,r,i,z$ CFHT images on 2019 Jan 15, and a series of 58 SDSS-r' $120\,\rm s$ images  on 2019 January 24 with the 2.54 m Isaac Newton Telescope (INT) on La Palma, Spain clearly showed both the previously reported tail and a new short dust tail subsequently reported by E. Jehin et al. \citep{CBET4606}.

\begin{figure}[!ht]
\centering
\includegraphics[width=8cm]{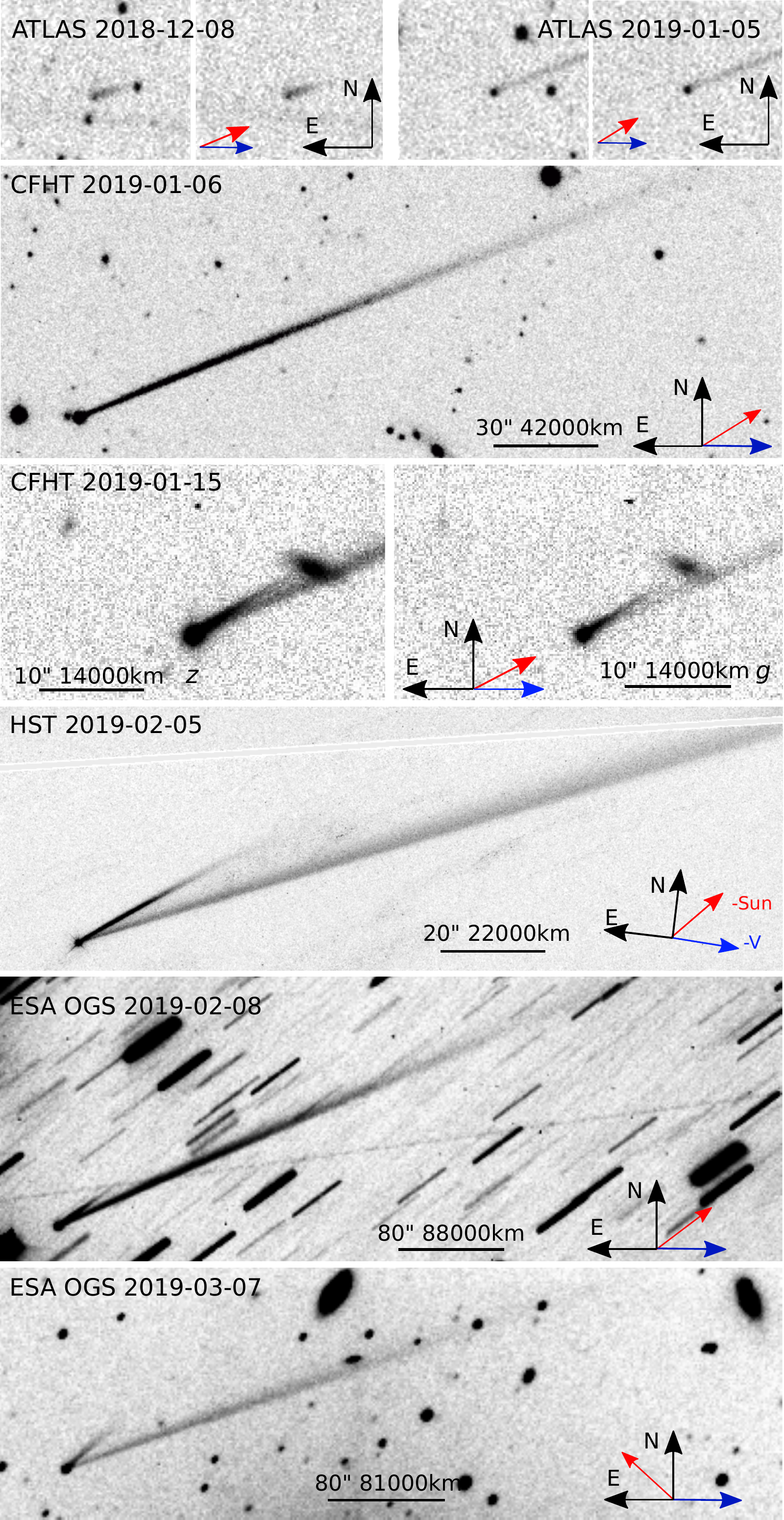}
\vspace{-0.3cm}
\caption{\label{fig:images}\small
Evolution of the (6478) Gault dust tail from the initial discovery in the ATLAS data. The tail on 2019 January 6 obtained with the CFHT 3.6 m telescope was 310$''$ (4.3$\times$10$^5$ km) long. The arrows indicate \ed{the orientation of the field and} the anti-solar direction and the negative of the object's velocity.  \ed{The images are individually
    adjusted on a negative logarithmic scale.} 
}
\vspace{-0.4cm}
\end{figure}

We were allocated three one-orbit observations with Hubble Space Telescope (HST) (program GO/DD-15678) to study the  evolution and morphology of the dust trail at high resolution and to search for possible fragments, with the goal of identifying the cause of the mass loss from the competing scenarios of sublimation, impact disruption, or YORP spin--up.  The first visit for this program was executed on 2019 February 5, yielding five dithered images of $380\,\rm s$ duration through the F350LP filter (WFC3/UVIS);  Fig.~\ref{fig:images} shows the stacked composite.

To assess Gault's rotation period, photometric data were obtained with the 1 m European Space Agency Optical Ground Station (OGS) at the Teide Observatory, Tenerife, on 2019 February 8.     One hundred fifty exposures of $90\,\rm s$  were obtained between 00:30 and 04:30 UT with a signal to noise ratio $(S/N)\gtrsim 100$.  

Additional data were obtained on 2019 February 10 with the 2 m Himalayan Chandra Telescope (HCT) located at Hanle-Ladakh, yielding 76 $R$-band images of $120\,\rm s$, with mostly modest extinction of $\lesssim 0.5$ mag, but occasional highs of $\sim 2$ mag.

Finally, a 351 exposure, 5.6-hour sequence of SDSS-r' band images was obtained on 2019 February 18 with the 4.2 m William Herschel Telescope (WHT) on La Palma. 
Conditions were photometric, but the images suffered from seeing of $\sim 1.5\arcsec-2.0\arcsec$, and from non-uniform scattered light from the Moon, 29$^\circ$ away.

\vspace{-0.2cm}

\section{Dust Dynamical Model\label{sec:dust}}

\begin{figure*}[!ht]
\includegraphics[width=18cm]{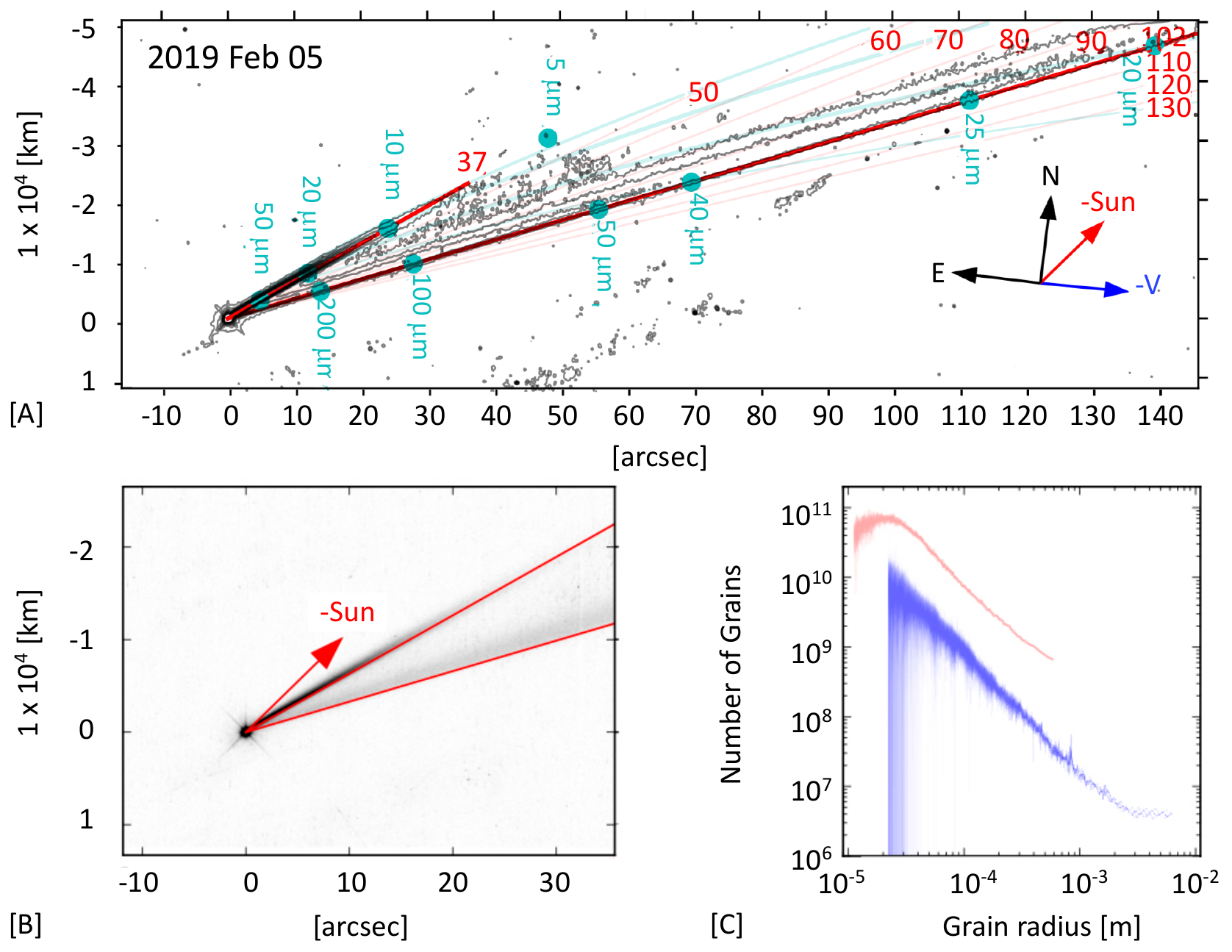}
\caption{\label{fig:FPfull}
\vspace{-0.6cm}
[A] \small Syndynes (blue) and synchrones (red, labeled in days before the observation date) for Gault \ed{HST data}, $t$=2019 February 5. The thicker synchrones mark the sharp onset of the dust emission ($t$--102 days) and the peak of the secondary emission ($t$--37 days). The corresponding grain radii are marked in blue. The orientation of the field and the antisolar and anti-velocity vectors are shown. 
[B] Close-up view of [A]. The two red lines are the synchrones corresponding to emissions at $t$--102 and $t$--37 days. The arrow marks the antisolar direction, the direction toward which dust would be drifting if emitted at the time of the observations. The lack of dust between the 37-day synchrone and the arrow indicates that the activity essentially stopped. \ed{No diffuse coma nor fragments are visible around the nucleus.}
[C] Distribution of the grain sizes around the 102- (blue) and 37-day (red) synchrones. 
}
\end{figure*}

A Finson-Probstein \citep[FP;][]{finson1968,farnham1996} dust analysis  calculates the trajectories of  dust grains of different sizes, parameterized by $\beta$ (the ratio of  radiation force and  solar gravity) ejected from the asteroid's surface at different times, $t$, as acted upon by solar gravity and solar radiation pressure.  We used the FP approach to compare synchrones (loci of the particles emitted at the same time $t$) and syndynes (curves joining particles with the same $\beta$)  to HST images of Gault's dust environment (see Fig.~\ref{fig:FPfull}).
$\beta$  is related to the size of dust grains by 
\begin{equation}
\beta = 5.740 \times 10^{-4} \times \frac{ Q_{\rm pr}}{ \rho a} ,
\label{Eq:beta}
\end{equation} 
\noindent 
where $Q_{\rm pr}$ is a radiation-pressure efficiency coefficient ($\sim$1--2 for rocky and icy material), $\rho$ is the  density, and $a$ is the  grain size. For $\rho = 3000$~kg~m$^{-3}$ and $Q_{pr} = 1.05$, Eq.~\ref{Eq:beta} yields $ a = 2\times 10^{-7}  \beta^{-1}$~[m].  

The dust emission started abruptly at $t-102$~days (2018 October 26) before the HST observations (t=0), \ed{in agreement with the ZTF observations of \cite{ATEL12450}}. No dust is observed on older synchrones, and the boundary of the \ed{trail} matches the synchrone perfectly, indicating a sharply defined event \ed{(i.e. shorter than the resolution of the FP modelling, 1~day)}and a broad distribution of large dust grains, with $\beta < 0.01 $, or $a > 20$~$\mu$m. Smaller dust grains were pushed out of the field of view. The dust emission peaks at $t-100$~days (2018 October 28), then decreases to almost no dust at $t-85$~days (2018 November 12). However, while the onset of the activity is sharply delimited by the $t=102$~day synchrone, the turnoff is not precisely aligned with the $t-85$~day synchrone. The streamer has a fairly constant width, while the distance between the synchrones increases radially, indicative of a small ejection velocity. 

\ed{The data show that a} second episode of dust emission took place around day $t-37$ (2018 December 30, \ed{$\pm$ 1 day}). This episode was  short and peaked, with a FWHM of 1~day; \ed{however, \cite{ATEL12450}  suggest 
that the event developed over several days.}  While this \ed{trail} is much shorter than the first, it shows smaller dust ($a \gtrsim 5$~$\mu$m), as the grains have had less time to be removed by radiation pressure. No dust is visible on the synchrones ranging between the second peak and the time of the observation.

The dust synchrone orientations for the ATLAS observations were PA=288$^{\circ}$ for 2018 December 8 and PA=291$^{\circ}$ for 2019 January 5 (in excellent agreement with the \ed{ATLAS} reported directions PA=290$^{\circ}$ and PA=291$^{\circ}$, respectively). The computed trail lengths out to $\beta = 0.02$ were $32\arcsec$ and $120\arcsec$, versus $30\arcsec$ and $135\arcsec$ reported from the observations. Thus the dust trail reported from the ATLAS observations matches the trail corresponding to the first release of dust on 2018 October 28.

The second dust release on 2018 December 30 was present in the 2019 January ATLAS and CFHT observations, but not identified as a trail. Its presence explains the reported $>1$ magnitude excess in the seeing disk.

The area around the nucleus is devoid of dust, indicating that the dust was released with tiny initial velocity.  No fragments are visible down to $\sim 50$~m radius.

Profiles were extracted along the synchrones over both trails.  The value of $t$ is obtained from the PA of the synchrone.  The linear position along these profiles were converted into $\beta$ using the FP model, and into grain radius using Eq.~\ref{Eq:beta}. The flux of a grain of radius $a$ is estimated as 
\begin{equation}
     f = 10^{ -0.4 (M_\sun - {\rm ZP} )} p \left(\frac{a}{r \Delta}\right)^2,
\end{equation}
where $f$ is the flux in CCD adu pixel$^{-1}$ s$^{-1}$, $M_\sun$ is the absolute magnitude of the sun in the filter, ZP=26.817 the photometric zero point\footnote{http://www.stsci.edu/hst/wfc3/analysis/uvis\_zpts/uvis1\_infinite}, $p=0.2$ is the dust albedo, $a$ the radius, and $r$ and $\Delta$ are the helio- and geocentric distances (all in astronomical units). 

These distributions show a dispersion, suggesting that the conversion from the position along the synchrone to $\beta$ and $a$ is not perfect. The same exercise was repeated, this time fixing the value of $t$ to the peaks of the streamers at 102 and 37~days. The resulting sharper distributions are shown in Fig.~\ref{fig:FPfull}.c. This suggests that the azimuthal spread of the streamer is dominated by an initial velocity rather than by a spread in emission time, independently supporting the earlier conclusion, based on the rectangular shape of the first streamer, that the azimuthal spread was caused by a distribution of initial velocities rather than purely by dust dynamics. The direction of the emission velocity is not known, but the improvement of the profiles using a constant $\beta$ suggests that a velocity perpendicular to the streamer is a good approximation. This  neglects a component perpendicular to the plane of the sky, which cannot be estimated. The spread of the streamer measured perpendicularly to its length ($l=5700$~km) and the age of the streamer ($t\sim 102$~days) give a lower limit of $v_e = 0.7$~m~s$^{-1}$ for the maximum emission velocity.
 
A power law of $f(a)~{\rm d}a = C ~ a^n ~{\rm d}a$ was fitted to grain size distribution profiles, resulting in indices $n=-1.70 \pm 0.08$ and $-1.64 \pm 0.01$ over the linear ranges for the first and second release events, respectively.  The value of $a$ and the number of grains of that size are affected by the assumed values of the density $\rho$ and the albedo $p$, but the index of the power-law is not.

In comparison, the streamers of activated asteroid 311P/2013~P5 (PANSTARRS) had a power-law index $-1.0$ \citep{hainaut2014}. Traditional sublimating comets have indexes in the $-4$ to $0.0$ range 
\citep{sekanina1980, fulle2000, moreno2016, moreno2017}.  

The number of grains in each pixel can also be used to estimate the mass of dust in the streamers: with the same assumptions ($p=0.2$, $\rho=3000$~kg~m$^{-3}$), this results in $m = 7 \times 10^9$ and $4 \times 10^7$~kg for the first and second streamers, respectively, integrating over $a$ from 30~$\mu$m to 2~mm. \ed{These are lower limits, as the mass of the streamer is dominated by the large particles, which the radiation pressure has not dispersed much and whose photometric contribution is small. The smaller particles, while more numerous, do not contribute much to the total mass \citep[see][Figure 10 for a quantitative discussion]{hainaut2012}}. To put these values in perspective, they correspond to spheres of 82 and 14~m radii,  small compared to the bulk of the body. The mass in the two main trails of P/2013~P5, another active asteroid which presented similar morphology, was estimated using a similar method to $3 \times 10^6$~kg and $3 \times 10^7$~kg \citep{hainaut2014}. P/2012~A2, also an active asteroid, presented a trail with a different morphology whose mass was estimated to be $8 \times 10^8$~kg \citep{hainaut2012}. 

\section{Rotation Period\label{sec:rotation}}
Using the OGS, HCT, and WHT data, we performed an analysis of Gault's light curve to determine the rotation period. The presence of a dust coma necessitated a small $2\arcsec$ aperture, which made our analysis susceptible to seeing variations.  Nevertheless, after linear detrending all three data sets showed a $\sim 1\,\rm h$ signature in the spectral analysis, in agreement with a two-peaked $\sim 2\,\rm h$ rotation period.
This is close to the critical breakup limit of a strengthless rubble pile ($\sim$3.3 hr for a cometary object, and 1.9 hr for a asteroid, with an absolute magnitude $V=14.4$) and at the observed 2 hr spin limit of asteroids \citep[e.g.][]{Pravec2002}.

\begin{figure}[!ht]
\centering
\includegraphics[width=8cm]{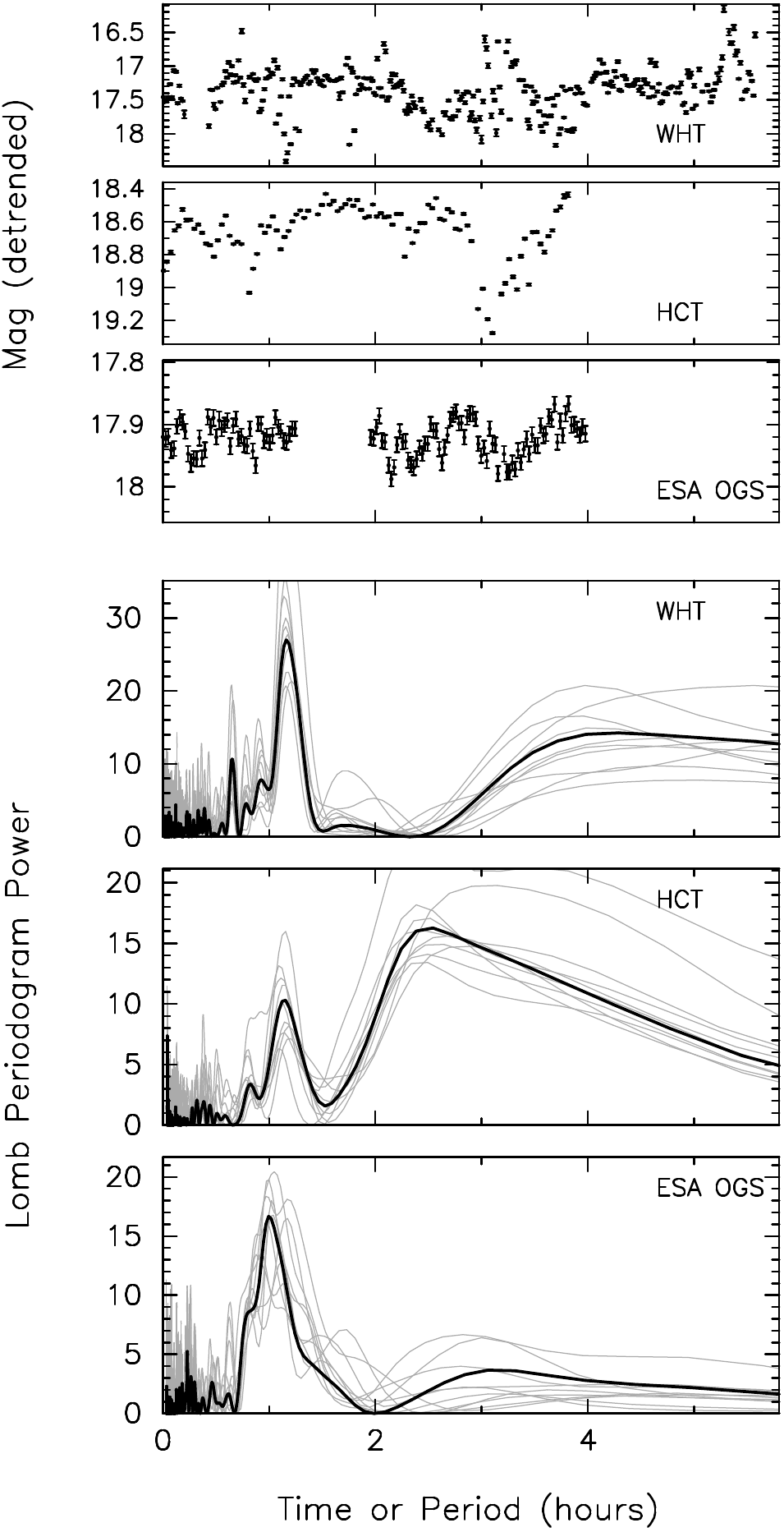}
\vspace{-0.3cm}
\caption{\label{fig:rot}
\small Top three panels: magnitudes versus time for  WHT, HCT, and OGS, after linear detrending.  Bottom three panels: The Lomb-Scargle periodogram of spectral power versus period for these data, with all sets showing a $\sim 1\,\rm h$ spectral peak, corresponding to a $\sim 2\,\rm h$ rotation. The gray curves are spectrograms for  resamplings of the data, indicating that the spectral signature is robust within each series, despite the noisy nature of the data.}
\end{figure}

Figure \ref{fig:rot} shows the data sets in the top three panels and the Lomb-Scargle spectral power \citep{Lomb1976} in the bottom panels, with Monte Carlo resamplings. \ed{The spectral peaks have formal significances of $p=1\times10^{-9}$ (WHT, 1.14 hr period); $p=6\times10^{-3}$ (HCT, for the shorter 1.16 hr period); and  $p=2\times10^{-5}$ (OGS, 0.97 hr period).  However, phasing and smoothing the data does not reveal any obvious light curve, suggesting that the periodic signal is buried in aperiodic, non-Gaussian noise, and has low amplitude (perhaps $\lesssim 0.05\,\rm mag$). A small amplitude light curve is expected if the dust coma contributes most of the flux \citep[e.g.][]{hsieh2011}, a supposition supported by the $\sim 1\,\rm mag$ brightening noted above.}

Repeating the test with a two-Fourier-component Analysis of Variance (ANOVA) analysis using the {\sc Peranso} package \citep{paunzen2016} also finds a two-hour period. A joint analysis of the temporally proximate WHT and ESA sets in {\sc Peranso} also shows a 2-hour rotation, although the signal in the OGS set must be scaled up, as might be required of a signal masked by dust in the OGS' large $1.4 \arcsec$ pixels.  The INT data, spanning 3 hr, also had a broad but insignificant 2 hr rotation. \ed{Although the absence of a visible light curve precludes definitive conclusions about the period, the presence of a 2 hr rotational signature in three distinct data sets, under two methods of analysis, with robustness under Monte Carlo resampling, is persuasive.}

An analysis of the 323 sparse observations found in  ATLAS  from 2016 to 2018 did not detect a signal; however, ATLAS cannot rule out variations with an amplitude of $\leq 0.05$ mag, \ed{in accord with the low amplitude inferred from our optical data.  }

\section{Dynamical Properties \label{sec:dynamics}}

Gault is a dynamical member of the Phocaea family \citep{nesvorny2015_pdsastfam}, which has been estimated to be $(1.2\pm0.3)$~Gyr old \citep{milani2017_astfamages}. The Phocaea region is a high-eccentricity, high-inclination portion of the inner asteroid belt  dominated by S-type asteroids \citep{carvano2001_hungaria_phocaea} and is bounded by the 3:1 and 4:1 mean-motion resonances (MMRs) with Jupiter and the $\nu_{5}$, $\nu_{6}$, and $\nu_{16}$ secular resonances.

\ed{Using the Hierarchical Clustering Method \citep{zappala1990_hcm,zappala1994_hcm},
we find that Gault also becomes dynamically linked with the overlapping \ed{low-albedo} Tamara family at a cut-off velocity of 113~m~s$^{-1}$, well below the threshold of 350~m~s$^{-1}$ identified for the family \citep{novakovic2017_phocaeafamily}.  Gault's albedo is currently unknown, however, and so its physical association with this family is uncertain.}

To assess whether Gault is  native to its current location in orbital element space or is a recent interloper like a \ed{dynamically evolved} Jupiter-family comet, we analyze its long-term dynamical stability.  We generate 10 dynamical clones of Gault with Gaussian-distributed orbital elements centered on the object's osculating orbital elements on 2019 February 7, \ed{using} $\sigma$ values equivalent to the orbital uncertainties  ($\sigma_a=9\times10^{-9}$~au, $\sigma_e=4\times10^{-8}$, $\sigma_i=5\times10^{-6}$~degrees).  \ed{Following the method of \citet{hsieh2012_scheila}}, we then perform forward integrations for Gault and its clones for 100~Myr \citep[substantially longer than typical dynamical lifetimes for short-period comets;][]{levison1994_spcdynamics}, using the Bulirsch-St{\" o}er integrator in the Mercury $N$-body integrator  \citep{chambers1999_mercury}.  

\begin{figure*}[!ht]

\centering
\includegraphics[width=17cm]{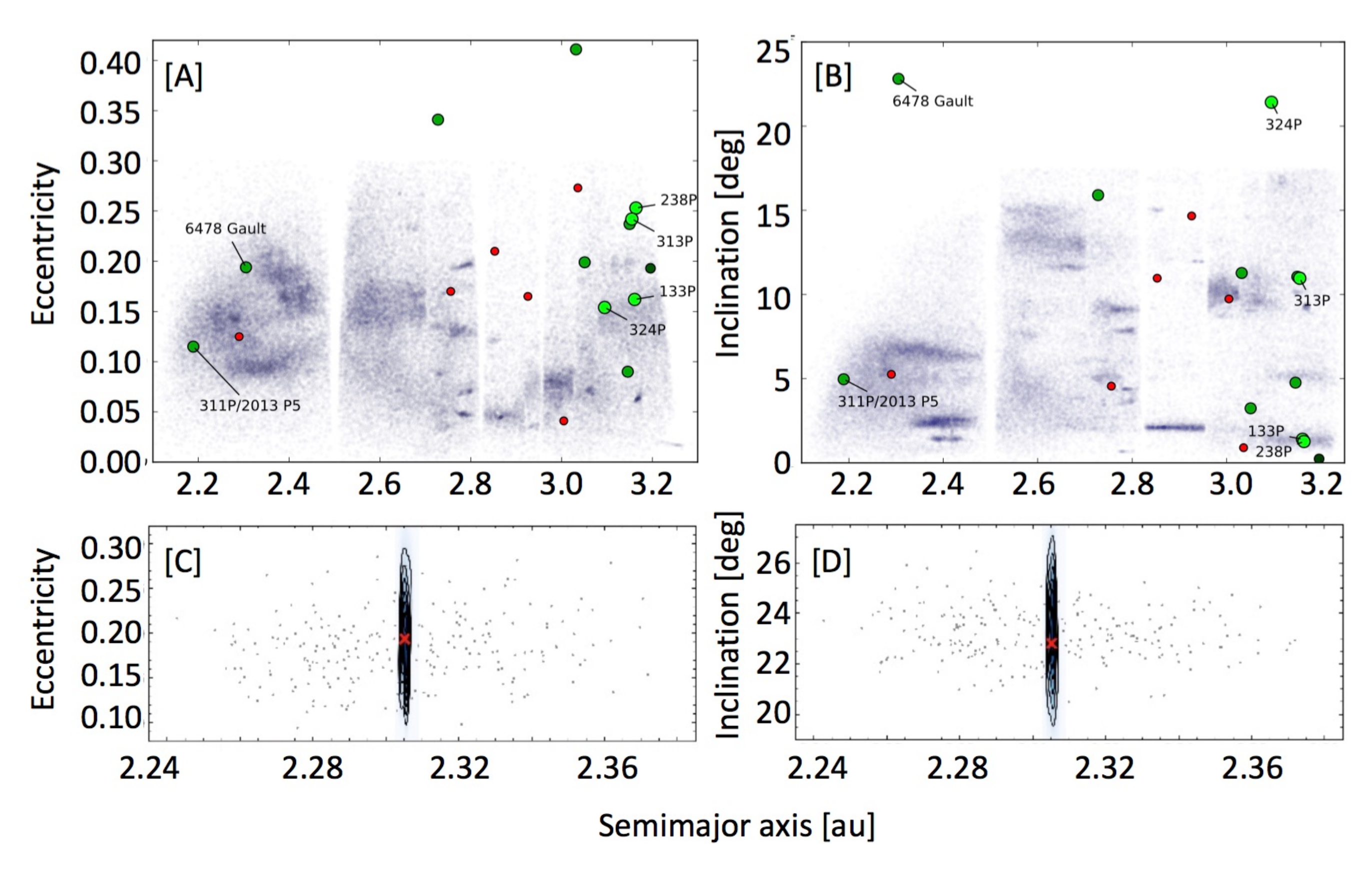}
\vspace{-0.3cm}
\caption{\small [A,B] Orbital elements. Gault and some other activated asteroids and main-belt comets are identified. Red dots correspond to objects likely to have been activated via collision. [C,D] Contour plots (black lines) of intermediate orbital elements in time steps of $10^4$ years in osculating semi-major axis vs.\ osculating eccentricity space and inclination space for 100~Myr forward integrations of Gault and 10 dynamical clones with orbital elements within 1~$\sigma$ of Gault's catalogued osculating orbital elements. The original osculating orbital elements of Gault and its clones are marked with a red cross in each panel.  The current osculating orbital elements of the members of the Tamara family (small gray dots) are also shown for reference.
\label{fig:orbital_evolution}}
\end{figure*}

\ed{For broader context, we also perform the same analysis for two sets of 50 clones with $\sigma$ values 10 and 100 times larger than Gault's orbital element uncertainties.}
Only one object in the set of clones created using the largest $\sigma$ values is ejected from the solar system (defined as reaching $a>100$~au) during our integrations. All other particles in all sets of clones remain stable for the 100~Myr integration period, with minimal deviations in semimajor axis ($\Delta a<0.01$~au) and osculating elements staying largely within the confines of the Tamara family (Figure~\ref{fig:orbital_evolution}).  These results indicate that Gault is unlikely to be a recently implanted interloper.   

\section{Discussion}

The presence of a $\sim2$ hr signature in the three data sets identifies Gault as a superfast rotator near or at the limit of a body with some internal cohesion \citep{holsapple2007, chang2019}. Hence dust emission is strongly suggestive of a rotation-induced event due to the YORP effect, as the object is spun-up by re-radiation forces until the apparent surface gravity is zero, triggering disruption or landslide events \citep[e.g.][]{scheeres2015}, releasing near-zero-velocity debris that is swept away by radiation pressure. Sudden and brief landslides are in accord with the abrupt dust releases described in \S\ref{sec:dust}. Because of the large mass of material released, it is likely that these landslides were significant, and that the equatorial velocity of the object is very close to the liberation velocity, i.e., that the surface material is tenuously held to the surface, with a proclivity to rearrange itself. The observed dust velocity of up to $\sim0.7\,\rm ms^{-1}$ is in accord with the $\sim 2\,\rm ms^{-1}$ surface velocity of a $4\,\rm km$ object rotating with a 2 hour period -- i.e., there is no unexplained source of energy. We might see more activations in the future.  The low amplitude of the light curve may be explained if the body has already been rotationally reformed to a nearly round, top-like shape, akin to asteroids Ryugu and Bennu.  

It is even possible that we are catching Gault in the process of episodic landslide transformation from a Maclaurin spheroid to a slower-rotating Jacobi ellipsoid, perhaps initiated by a collision. In such an event the period would slow down, and the light-curve amplitude should increase. This is consistent with the absence of rotation in the ATLAS data.  Further monitoring of the rotation curve before and after any future emission events is warranted.

The characteristic timescale for a YORP spin-up of a $4\,\rm km$ object is  $\sim 10^8$ years \citep{Bottke2006}, well within Gault's $\gg 10^8\,\rm yr$ stability time constraint of \S\ref{sec:dynamics}.

\section{Summary} 

Gault experienced two dust releases  occurring around 2018 October 28 and December 30, creating the observed streamers \citep{ATEL12450,CBET4606}. The October 28 streamer was observed by ATLAS and CFHT.
The width of the first streamer is best explained with a maximum emission velocity $v_e \sim 0.7$~m~s$^{-1}$ in the sky plane. These events were short, with  upper duration limits of ${\ll}15$~days for the first, and ${\ll}5$~days for the second.  The size distribution of the dust grains in the streamers follows a power law with an index $\sim -1.65$. The mass lost in the streamers is  $m \approx 7 \times 10^9$ and $4 \times 10^7$~kg, respectively.

Dynamical simulations show that Gault is dynamically stable and unlikely to have been recently implanted from elsewhere, ruling out a cometary origin.

The presence of a $\sim 2$ hour signature in  three data sets  identifies Gault as a superfast rotator that likely underwent a YORP-induced rotational disturbance.

\section{ACKNOWLEDGMENTS} 

We acknowledge the following supporting grants: KJM: NSF award AST1617015, and HST program GO/DD-15678 from NASA through STScI, operated by AURA under NASA contract NAS 5-26555;  HHH: NASA Solar System Workings grant 80NSSC17K0723;  AF: UK STFC grant ST/P0003094/1.

This work uses data from the ATLAS project, funded through NASA grants NN12AR55G, 80NSSC18K0284, and 80NSSC18K1575, with the IfA at the University of Hawai'i, and with  contributions from the Queen's University Belfast,  STScI, and the South African Astronomical Observatory.

The INT and WHT are operated on island of la Palma, by the Isaac Newton Group of Telescopes in the Spanish Observatorio del Roque de los Muchachos of the Instituto de Astrofisica de Canarias. 

We thank the staff of IAO, Hanle, and CREST, Hosakote, that made these observations possible. IAO and CREST are operated by the IIAP, Bangalore.
\ed{We also thank the director and staff of the CFHT, for enabling us to obtain observations on short notice.}

\vskip 0.5cm

\noindent{\it Note added in proof:} During the proofing stage of this paper, we learned of an independent paper by \cite{ye2019ApJL} using ZTF light curve data to reach a similar conclusion that Gault's activity is attributable to rotational disruption or a YORP-driven binary merger.

\clearpage

\bibliographystyle{aasjournal}

\input Gault-Paper1.bbl
\end{document}

%% file: ObsTable.tex

\begin{deluxetable*}{lcccccccccc}
\tablecaption{Observations\label{tab-obslog}}
\tablecolumns{11}
\tablehead{
\colhead{Telescope}       & 
\colhead{UT Date}         &  
\colhead{Exp$^{\S}$}      &
\colhead{\#Exp}           &
\colhead{Filter}          &
\colhead{Sky}             &
\colhead{Seeing}          & 
\colhead{$r^{\dag}$}      &
\colhead{$\Delta^{\dag}$} &
\colhead{$\alpha^{\dag}$} &
\colhead{TA$^{\dag}$} 
}
\startdata
\hline
CFHT MCam & 2019 Jan 6  &  60     & 6   & g,r,i     & clear        & 1.0  & 2.47 & 1.88 & 21.0 & 238 \\
CFHT MCam & 2019 Jan 8  &  180    & 6   & w         & clear        & 0.8  & 2.47 & 1.85 & 20.7 & 238 \\
CFHT MCam & 2019 Jan 15 &  120    & 17  & g,r,i,z   & clear        & 0.6  & 2.47 & 1.88 & 21.0 & 238 \\    
INT       & 2019 Jan 23 & 120     & 58  &       r   & clear        & 2.0  & 2.44 & 1.68 & 17.6 & 242 \\
HST WFC3  & 2019 Feb 05 & 380     & 5   & 360LP     & N/A          & N/A  & 2.41 & 1.54 & 13.7 & 245 \\
CFHT MCam & 2019 Feb 06 &  90-120 & 14  & g,r,i,z   & clear        & 1.2  & 2.46 & 1.76 & 19.4 & 240 \\    
ESA OGS   & 2019 Feb 08 &  90     & 150 &  Open     & clear        & 2.0  & 2.41 & 1.52 & 12.6 & 246 \\
HCT 2m    & 2019 Feb 10 & 120     &  76 &       R   & cloud        & 2.0  & 2.40 & 1.50 & 11.8 & 246 \\ 
WHT       & 2019 Feb 19 &  40     & 350 &       r   & clear, moon  & 1.7  & 2.38 & 1.44 &  8.7 & 249 \\
ESA OGS   & 2019 Mar 07 &  90     &   3 &  Open     & clear        & 2.0  & 2.35 & 1.39 & 7.2  & 253 \\
\hline
\enddata
\tablenotetext{$\S$}{Image exposure time (s)}
\tablenotetext{$\dag$}{Heliocentric and geocentric distance (au), phase angle (deg), and true anomaly (deg)}
\end{deluxetable*}